\documentclass{moriond}

\def\be{\begin{equation}}
\def\ee{\end{equation}}
\def\bea{\begin{eqnarray}}
\def\eea{\end{eqnarray}}

\newcommand{\Photo}{\includegraphics[height=35mm]{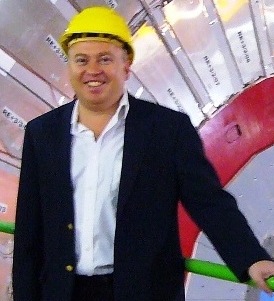}}

\begin{document}
\vspace*{4cm}
\title{MOVING BEYOND EFFECTIVE FIELD THEORY IN DARK MATTER SEARCHES AT COLLIDERS}

\author{GREG LANDSBERG }

\address{Department of Physics, Brown University, 182 Hope St, Providence, RI 02912, USA\bigskip}

\maketitle\abstracts{
In the past few years, the interest to collider searches for direct dark matter (DM) production has been growing exponentially. A variety of ``Mono-X" signatures have been considered, where X stands for a probe particle recoiling against DM particles, which allows for the event to be triggerable. So far, the analysis of these signatures has been largely carried out in the framework of effective field theory (EFT), which allows for a comparison of the collider searches with searches in direct detection experiments. Unfortunately, as it has been recently pointed out by a number of authors, the EFT approach has severe limitations and may result in drastically underestimated or overestimated reach. I'll discuss these limitations and the new ideas in interpreting the collider searches for DM.\vspace*{-6pt}}

\section{Introduction}

Effective field theory (EFT) has been an important tool to study various processes where a detailed description of the interaction and its carrier is either unknown or model-dependent. The EFT is used to parameterize our ignorance of the fine details of the process and has been successfully applied to a number of cases, including Fermi's model of muon decay and searches for compositeness. It is therefore logical that the original theoretical papers~\cite{DM1,DM2,DM3} that proposed the initial-state radiation (ISR) tagging to detect dark matter production (DM) at colliders, relied on the EFT description of the scattering process in order to allow for a comparison of the sensitivity of these searches with that for direct detection (DD) experiments. A classical example of such a collider process is production of a single jet recoiling against a pair of DM particles that escape the detection, resulting in a spectacular ``monojet" signature. Similar, ``monophoton" signature is also possible in the case of a photon ISR.

Unfortunately, as has been realized recently, the use of EFT in this particular case is subject of a number of explicit and implicit assumptions, and important constraints, which severely limit the applicability of the EFT approach, sometimes to the point when it becomes all but useless. In this particular application, the EFT often fails in all three possible ways:
\begin{itemize}
\item As an "E" --- not being effective in probing certain regions of parameter space;
\item As an "F" --- sometimes not even dealing with realistic fields; and
\item As a "T" --- not even holding as a viable theory.
\end{itemize}
The goal of these proceedings is to illustrate the limitations of the EFT approach and discuss more constructive ways of comparing the DM reach of collider experiments with that of the DD experiments, and potentially also with the reach of indirect detection experiments. Such a proper comparison would become particularly important if a significant excess in any of these experiments is seen.

\section{EFT formalism and assumptions}

\begin{figure}
\begin{minipage}{0.5\linewidth}
\centerline{\includegraphics[width=0.9\linewidth]{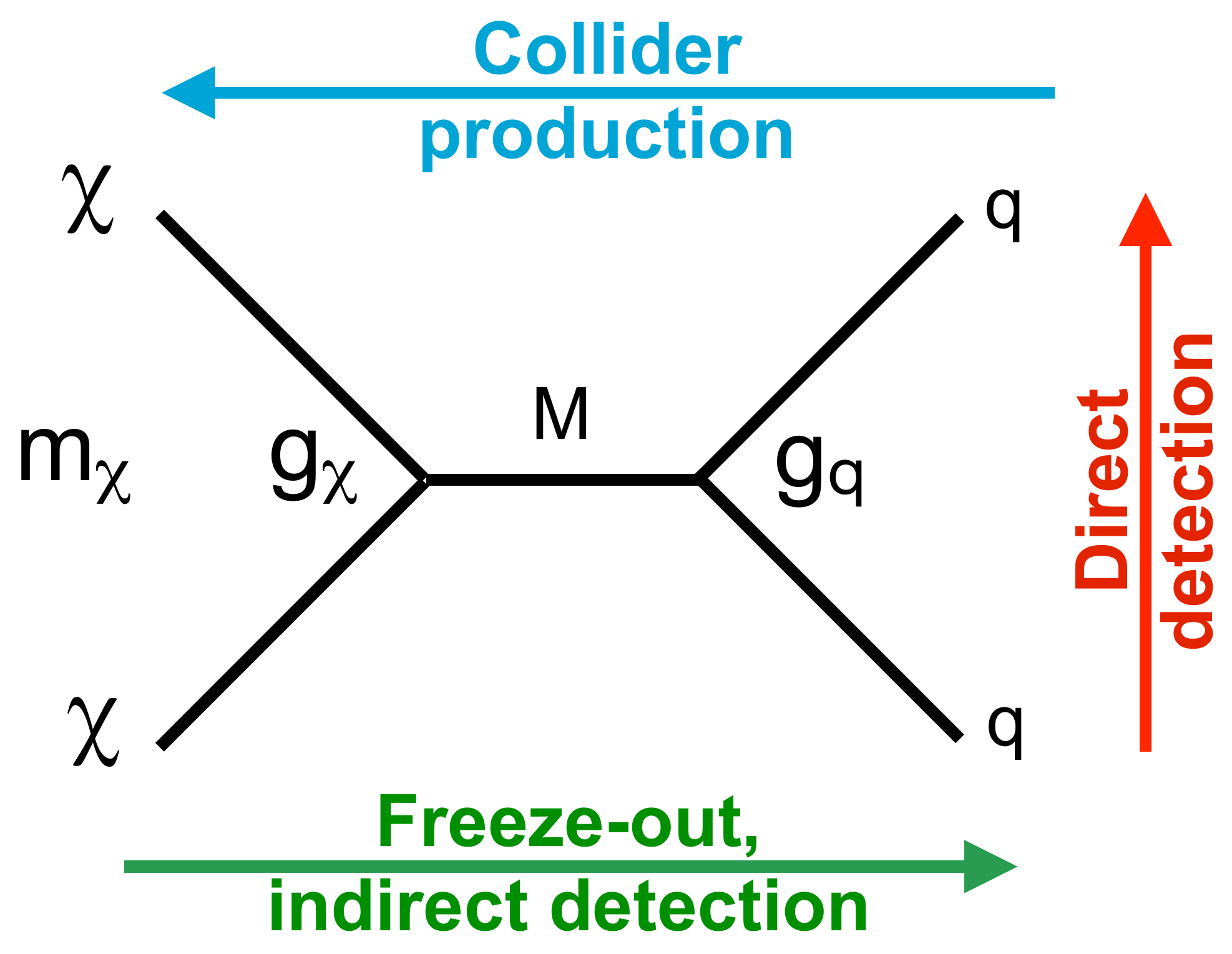}}
\end{minipage}
\hfill
\begin{minipage}{0.5\linewidth}
\centerline{\includegraphics[width=0.7\linewidth]{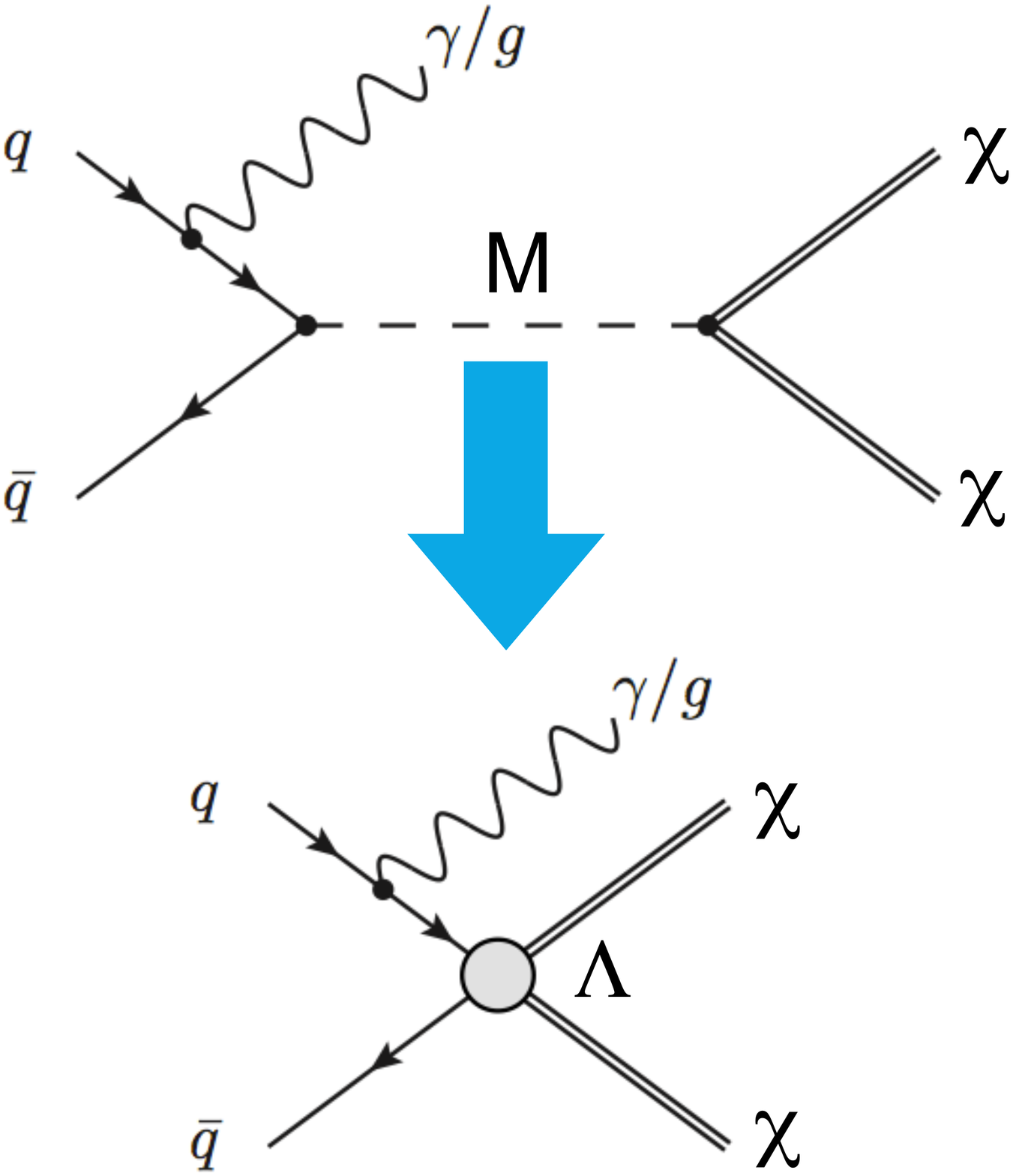}}
\end{minipage}
\vspace*{-12pt}
\caption{Left: Feynman diagram of dark matter interaction with quarks via exchange of a mediator. Right: ``contraction" of the s-channel mediator exchange diagram for the monojet or monophoton production into an EFT four-point interaction.}
\label{fig:Feynman}
\end{figure}

Collider experiments are capable of setting limits on production cross section of DM particles in ISR-triggered processes, e.g. production of monojets~\cite{CMS-monojet,ATLAS-monojet}. These limits only require theoretical calculations, which properly describe the ISR process. While next-to-leading-order calculations are available for many such processes, often leading-order precision with an extra jet emission included in the matrix elements, suffices, making it relatively easy to calculate collider cross sections. The real issue comes when collider limits are being translated into limits on DM-nucleon scattering cross section, which is the variable used by DD experiments to represent their results. Note that fundamentally the process responsible for pair production of DM particles at colliders is the same as for the DM-nucleon scattering, or annihilation of a pair of DM particles used in indirect detection experiments. Assuming that the former process is mediated via an s-channel exchange of a certain particle, which we will refer to as the ``mediator", the process is completely described by four parameters: the masses of the DM particle ($m$) and the mediator ($M$), and the two couplings of the mediator to quarks ($g_q$) and DM particles ($g_\chi$), see Fig.~\ref{fig:Feynman} (left). (A similar diagram can be drawn to describe collider DM pair production via a t-channel exchange of a mediator, with the caveat that in this case the mediator must be a colored particle.) In order to compare the s-channel collider process with the t-channel DM-nucleon scattering, we ``contract" the s-channel exchange in the EFT four-point interaction vertex, as shown in Fig.~\ref{fig:Feynman} (right), which then can be used to describe both. In order to perform this contraction we move from three fundamental parameters $M$, $g_q$, and $g_\chi$ to a single parameter $\Lambda$, the EFT cutoff, thus losing the full information about the underlying process, which is an inherent feature of the EFT approach.

One can now directly equate the amplitude squared of the s-channel exchange in the limit of a heavy mediator ($M^2 \gg q^2$ in the event) with the one from the effective four-point interaction, which for, e.g. a mediator with scalar couplings, yields: 
$$\left|\frac{ig_q g_\chi}{q^2 - M^2}(\bar q q)(\bar\chi\chi)\right|^2 \approx \left|\frac{-ig_q g_\chi}{M^2}(\bar q q)(\bar\chi\chi)\right|^2 = \left|\frac{1}{\Lambda^2}(\bar q q)(\bar\chi\chi)\right|^2,
$$ 
leading to a crucial expression: $\frac{1}{\Lambda^2} = \frac{g_q g_\chi}{M^2}$.

The EFT approach is strictly valid for $q^2 \ll M^2$, which implies (from the kinematics of the s-channel exchange) $M^2 > (2m)^2$. Furthermore, in order for theory to be calculable, each of the two mediator couplings has to be less than $\sqrt{4\pi}$. Combining these two inequalities with the expression for $\Lambda$, we obtain: $2m < M < \Lambda\sqrt{g_q g_\chi} < 4\pi\Lambda$, or $\Lambda > \frac{m}{2\pi}$, which leads to an important conclusion that the validity region of the EFT grows when one deals with light DM. Similar validity regions in case of non-scalar couplings can be found, e.g. in Ref.~\cite{Whiteson} The case of light DM is particularly important for colliders as the sensitivity of DD experiments to light DM is reduced due to low-momentum recoil, and since for very light DM ($m < 10$ GeV), the DD experiments will soon reach the solar neutrino floor. Nevertheless, it's important to keep in mind that the above inequality really corresponds to the case when all the EFT assumptions break down spectacularly, and actual validity region really corresponds to $\Lambda \gg \frac{m}{2\pi}$.

The most tricky scenario is the case of a light mediator, for which EFT certainly fails. This case was explicitly studied in one of the early phenomenological papers on collider searches~\cite{DM4}, with an explicit use of the s-channel exchange diagram instead of the EFT approach. In this case, collider searches offer an increased sensitivity to the DM production as they can produce light mediator on-shell, and hence the production cross section receives a resonant enhancement. However, the problem with the approach taken in Ref.~\cite{DM4} is that it treats the mediator width as a free parameter, whereas one can't do this, as the width of the mediator depends on the $\sqrt{g_q^2 + g_\chi^2}$, and if even one of the couplings approaches the $\sqrt{4\pi}$ limit, the width becomes comparable to the mass of the mediator, independent on how small the other coupling is. Since a single-resonance exchange description stops being physically reasonable for mediators that broad, this seemingly correct approach can still give incorrect comparison with the DD experiments~\cite{Oliver}.

\section{Beyond the EFT}

Given this situation, it is clear that EFT, while a convenient way to simplify the problem, has too many hidden caveats and simply does not allow for a fair comparison between the collider and DD experiments. The key to the proper comparison is to treat the problem as fundamentally four-dimensional and represent the reach of both the DD and collider experiments in various planes given by a pair of these parameters (e.g., $M$ and $m$), with the other two (in this case $g_q$ and $g_\chi$) being fixed to certain values, which can be scanned. In order to do this, one could use simplified models of DM, which assume certain type of couplings of the mediator to quarks and DM particles, e.g., vector or axial vector. Given that the number of such models is quite limited, one could rather easily span the relevant DM model space with just a handful of simplified models with s-channel or t-channel mediator exchange. Similar simplified model approach is successfully and broadly used in supersymmetry searches at the LHC. This is the approach advocated in the recent work~\cite{MSDM,MSDMA} coming from the two groups of experimentalists and theorists (the first one generally affiliated with the CMS experiment, whereas the second one -- with ATLAS). Both ATLAS and CMS are now transitioning to this approach to be used in the LHC Run 2.

\begin{figure}
\begin{minipage}{0.515\linewidth}
\centerline{\includegraphics[width=0.9\linewidth]{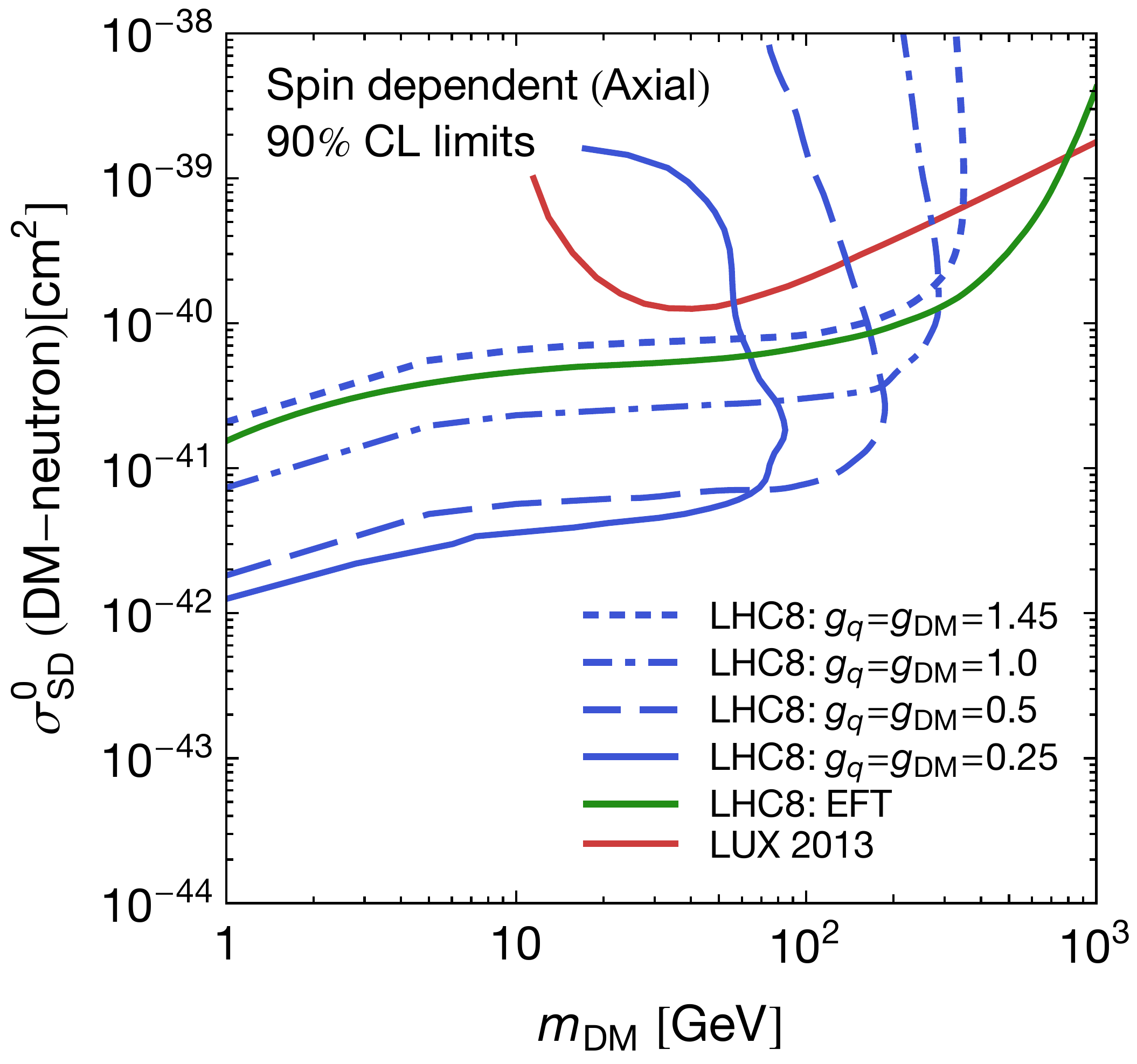}}
\end{minipage}
\hfill
\begin{minipage}{0.485\linewidth}
\centerline{\includegraphics[width=0.91\linewidth]{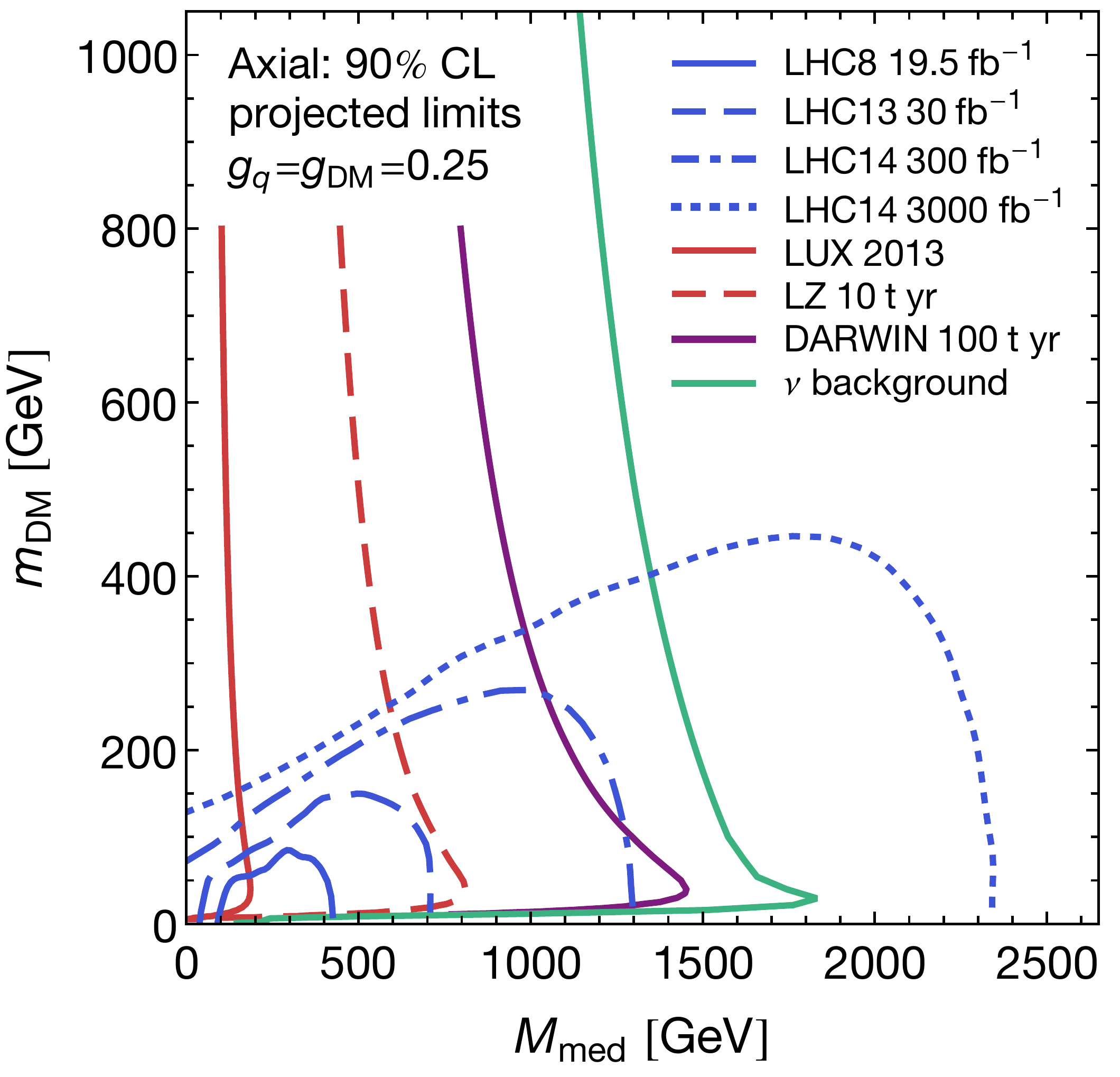}}
\end{minipage}
\hfill
\caption{Left: comparison of the EFT-based and simplified model limits on the DM-neutron scattering. Right: Comparison of the projected reach of the LHC and next generation of DD experiments. From Ref.~\protect\cite{MSDM}}
\label{fig:EFT}
\end{figure}

Figure~\ref{fig:EFT} (left) shows how the limits set using a simplified model with axial-vector couplings of the mediator to both DM particles and quarks compare with the limits from the EFT approach based on the CMS monojet analysis~\cite{CMS-monojet}, as well as with the limit from the LUX experiment~\cite{LUX} in the canonical plane of DM-nucleon scattering cross section vs. the DM particle mass. While for relatively large couplings $g_q = g_\chi = 1.45$ the EFT results are close to those from the simplified model calculations up to DM particle mass of about 300 GeV, for smaller values of couplings the EFT grossly underestimates the LHC reach for light DM and grossly overestimates it for relatively heavy DM. Figure~\ref{fig:EFT} (right) shows the projection of the CMS monojet analysis for LHC Run 2 and High-Luminosity LHC, as well as projected sensitivity of the next generation of DD experiments, in the more relevant plane of $M$ vs. $m$, for the case of axial-vector mediator couplings. One can see a nice complementarity between the reach of the two types of experiments, with LHC winning over DD experiments for the case of small couplings and relatively heavy mediators and in the case of very light DM particles (with the mass less than about 5 GeV), while DD experiments offering higher reach for rather heavy DM with the mass above 200-400 GeV. Similar comparison is possible with indirect detection experiments.

To conclude, the simplified model approach allows for a fair comparison of the DM reach of different types of experiments and provides a more clear and advantageous way to present the results of future collider searches.

\section*{Acknowledgments}

This work is partially supported by the DOE Award No. DE-SC0010010-003376. The author is grateful to the Imperial College, London CMS group for organizing and supporting the brainstorming workshop on dark matter at colliders, which sparkled this work and Ref.~\cite{MSDM}

\end{document}